\documentclass[preprint2]{aastex}

\shorttitle{The Log-normal Star Formation History of Galaxies}
\shortauthors{Gladders et al.}

\begin{document}


\title{The IMACS Cluster Building Survey. IV. The Log-normal Star Formation History of Galaxies}



\author{Michael D. Gladders\altaffilmark{}} \affil{The Department of
Astronomy and Astrophysics, and the Kavli Institute for Cosmological
Physics, The University of Chicago, 5640 S Ellis Ave., Chicago, IL 60637, USA}
\email{gladders@oddjob.uchicago.edu}

\author{Augustus Oemler\altaffilmark{} \& Alan Dressler\altaffilmark{}}
\affil{The Observatories of the Carnegie Insitution of Science, 813 Santa Barbara St., Pasadena, CA 91101}

\author{Bianca Poggianti\altaffilmark{} \& Benedetta Vulcani\altaffilmark{}}
\affil{INAF-Osservatorio Astronomico di Padova, vicolo dell'Osservatorio 5, 35122 Padova, Italy}

\and

\author{Louis Abramson \altaffilmark{1}}
\affil{Astronomy and Astrophysics, and the Kavli Institute for Cosmological
Physics, The University of Chicago, 5640 S Ellis Ave., Chicago, IL 60637, USA}


\altaffiltext{1}{present address: Brinson Fellow at The Observatories of the Carnegie Insitution of Science, 813 Santa Barbara St., Pasadena, CA 91101}


\begin{abstract}
We present here a simple model for the star formation history of
galaxies that is successful in describing both the star formation rate
density over cosmic time, as well as the distribution of specific star
formation rates of galaxies at the current epoch, and the evolution of
this quantity in galaxy populations to a redshift of z=1.  We show
first that the cosmic star formation rate density is remarkably well
described by a simple log-normal in time. We next postulate that this
functional form for the ensemble is also a reasonable description for
the star formation histories of individual galaxies. Using the
measured specific star formation rates for galaxies at z$\sim$0 from
Paper III in this series, we then construct a realisation of a universe
populated by such galaxies in which the parameters of the log-normal
star formation history of each galaxy are adjusted to match the
specific star formation rates at z$\sim$0 as well as fitting, in
ensemble, the cosmic star formation rate density from z=0 to z=8. This
model predicts, with striking fidelity, the distribution of specific
star formation rates in mass-limited galaxy samples to z=1; this match
is not achieved by other models with a different functional form for
the star formation histories of individual galaxies, but with the same
number of degrees of freedom, suggesting that the log-normal form is
well matched to the likely actual histories of individual galaxies.
We also impose the specific star formation rate versus mass
distributions at higher redshifts from Paper III as constraints on the
model, and show that, as previously suggested, some galaxies in the
field, particularly low mass galaxies, are quite young at intermediate
redshifts. As emphasised in Paper III, starbursts are insufficient to
explain the enhanced specific star formation rates in intermediate
redshift galaxies; we show here that a model using only smoothly
varying log-normal star formation histories for galaxies, which allows
for some fraction of the population to have peak star formation at
late times, does however fully explain the observations.  Finally, we
show that this model, constrained in detail only at redshifts $z<1$,
also produces the main sequence of star-formation observed at
$1.5<z<2.5$, again suggesting that the log-normal star formation
histories are a close approximation to the actual histories of typical
galaxies.

\end{abstract}


\section{Introduction}

Simple analytic models for the star formation histories (SFHs) of
galaxies have been explored for decades, and a number of basic SFHs
have found common usage in a variety of analyses. These SFHs range
from utilitarian models such as continous star formation or true
simple stellar populations (i.e., a delta-function burst), through to
more complex histories such as truncated continous star formation
\citep{lar78}, exponentially declining star formation \citep{sea73},
and the delayed exponential SFH \citep[e.g.,][]{gav02} as first
proposed by \cite{san86}.

Of course, all such simple models of entire galaxies are considered to
be time-averaged representations of a multitude of star formation
events occuring in individual star forming regions on timescales much
smaller than the dynamical timescale of a typical galaxy.  In detail
the SFH of a galaxy is almost certainly not easily described by a
simple function, but rather as the accumulation of a large number of
brief star formation episodes associated to some extent with the
merger history of that galaxy and its parent dark matter halo. An
appreciation of the importance of the effect of secondary bursts on
various observables has been present in the literature for decades
\citep[c.f.,][]{lar78}; secondary bursts are particularly
critical when predicting the observed properties of early type
galaxies, where bursts involving only a few percent of the galaxies
total stellar mass can significantly alter some spectroscopic and
photometric observables \citep[e.g.,][]{kav09}.

Paper III of this series \citep{gus12} presents a detailed look at the
measured star formation rates in field galaxy populations from z=0 to
z=1, and notes in particular that the observed evolution of the
specific star formation rate (sSFR; the star formation rate per unit
stellar mass) cannot be explained by simple models in which galaxies
are coeval populations, and furthermore that starbursts cannot modify
the sSFR distribution in such a model to match the observed values.
The data suggest that some galaxies must form the bulk of their stars
reasonably quickly at late times, from which we conclude that any
successful analytic model for the star formation histories of
individual galaxies must control both the onset time and timescale of
star formation; single parameter models such as exponentially
declining SFHs are insufficient.

In this paper we present a treatment of galaxy SFHs in which we use a
fit to the cosmic star formation rate density (SFRD) as the basis for
a functional form for the star formation history of typical
galaxies. The SFRD is a measure of star formation as a function of
lookback time and as such $is$ the SFH of the hypothetical average
galaxy. We explore the ability of aggregate populations of this simple
SFH to match fundamental observables such as the local and
intermediate redshift distribution of sSFRs in galaxies. The sSFR
distribution encodes the mass (i.e., the integrated SFH) and the
current star formation rate, and the relationship between these. We
motivate the functional form of the SFH - simply a log-normal in time
- in \S2, and provide simple fits to the data from Paper III using
only the SFRD and the z=0 sSFR distribution.

In \S3 we attempt to connect a suite of log-normal histories to
individual galaxies and add additional contraints from the measured
sSFR distribution of galaxies out to z=1. In \S4 we discuss the
implications of this modeling in the context of the conclusions of
Paper III, noting that this model succesfully produces some young
galaxies at late times without starbursts. In \S5 we discuss the
broader implications of the apparent efficacy of the log-normal SFHs. 
We summarize our main results and suggest some future tests of this model in \S6.
 
\section{The Log-Normal SFH} The ubiquity of the log-normal
distribution in nature is well known \citep[e.g.,][]{lim01}. Skewed
distributions such as this will arise with physical processes where
negative values are not permitted. Star formation, like many other
`growth' processes, is such a process; negative values of star
formation do not occur. It is thus reasonable to suppose that the SFH
of a typical galaxy might appear log-normal. Further, the history of
star formation in the universe - the star formation rate density
(SFRD) - is strongly skewed in time \citep[e.g.,][]{bou09} and is
at least reminiscent of a log-normal distribution, with a rapid rise at
early times, and long slow decline at late times. The log-normal
distribution is also a two-paramater function, and so at least in
principle can provide sufficient flexibility to match the sSFR
evolution highlighted in Paper III. This fact, the shape of the SFRD,
and the apparent ubiquity of log-normal distributions in nature
motivates the current analysis.

\subsection{The Log-normal Cosmic Star Formation Rate Density}

Current data \citep[see for example, the compilation in][]{cuc12}
clearly show both a rise and fall in the SFRD over cosmic time. More
specifically, the rise in the SFRD is fast at early times, with a long
continuing but declining tail of star formation to the present
epoch. Ignoring mergers, the shape of the SFRD should represent the
shape of the SFH of the mean galaxy in the Universe. Of course, the
actual SFH of any recognized single galaxy at z=0 will likely not fit
this description, not least because a typical z=0 galaxy is thought to represent
the merger of a number of smaller systems over cosmic time.

Putting aside this complexity for the moment, we note that the
presence of both a rise and fall in the SFRD also indicates that
simple SFH models which decline from intially high star formation
rates (i.e., exponentially declining models - referred to as $\tau$ models -
commonly used in modeling the spectrophotometric properties of
galaxies at later times) cannot easily describe the overall SFRD
evolution ; a delay time $must$ be allowed - such as in the delayed
exponential SFH of \cite{san86} - in order to even approximate
current data.  The rise in the SFRD from high redshift can be produced
by an ensemble of simple $\tau$ models if a second paramater - the
time at which such a SFH `turns on' - is allowed and is appropriately
distributed at early times. 

Recent analysis of the photometric properties of distant galaxies also
argue for a star formation history {\it in individual galaxies} that
is smoothly rising at early times \citep{mar10}. They found that a
star formation history of SFR$=Ae^{-(t-t_o)/\tau}$, a so-called
inverted-$\tau$ model, produced a better fit to the photometric
properties of a sample of $z\sim2$ actively star forming galaxies than
did the more typically used declining-$\tau$ models. A similar
result, for galaxies with a fixed comoving number density of
$2\times10^{-4}$ Mpc$^{-3}$, was found to hold from z=8 z to z=3 by
\cite{pap11}; \cite{red12} find that rising SFHs at $7>z>2$ are the
preferred history for a sample of galaxies observed at $z\sim2$.
Maraston et al. also note that the results of \cite{cim08} from a
study of $z\sim1.6$ elliptical galaxies also argue for an early star
formation history that is rising in time. Additionally, some
studies suggest that the sSFR of galaxies is flat beyond the peak of
the SFRD at $z\sim2$ out to at least $z\sim7$
\citep[e.g.,][]{sta09,gon10} though see \cite{deb12} for an alternate
view; one simple interpretation of such a flattening is that
individual galaxies at these redshifts have star formation histories
with a star formation rate smoothly increasing to later times. This
overall picture of rising SFHs in individual galaxies at high redshift
is also found in recent cosmological hydrodynamic simulations \citep{fin11,jaa12}.

Maraston et al. point out that the remarkably ubiquitous use of
declining $\tau$ models in the literature over the past few decades is
not well justified, and has become thoroughly divorced from the
original aim of these models of measuring the age of early-type
galaxies at z$\sim$0. It thus seems reasonable to seek a new simple
model which captures both the rising early-time and declining
late-time behaviour of galaxy SFHs.

In  seeking a functional  form to describe  the SFRD  over cosmic
time, we  have explored  a number of  possible descriptions,  and have
found  that a  remarkably simple  one ---  a log-normal in time --- works
extremely well.  We describe the scale free  log-normal distribution of
the star formation rate as

\begin{equation}
  SFR(t,t_0,\tau)=\frac{1}{t\sqrt{2\pi\tau^2}}e^{-\frac{(\ln t-t_0)^2}{2\tau^2}},   
\end{equation}


\noindent where $t$ is the elapsed time since the Big Bang, $t_0$ is
the logarithmic delay time, and $\tau$ sets the rise and decay
timescale. One great advantage of such a form over some other models -
such as the inverted-$\tau$ model of \cite{mar10} is that it naturally
subsumes both a rising and falling star formation history at different
times; simple $\tau$ models - inverted or otherwise - are aphysical,
whereas a log-normal in time could arguably describe the star
formation history of a galaxy (or at least a galaxy's components) at
all times. A further advantage is that the delay time is de-coupled
from the width of the star formation history, unlike the delayed
exponential model of \cite{san86}. \cite{beh12} have recently
suggested two other functional forms - a double power law, or a hybrid
exponential+powerlaw - which we will explore further below.

Figure 1 shows a fit to the SFRD with a single log-normal. This simple
functional form produces an reasonable fit, with a $\chi^2$ of 2.1.
It is worth re-empahsizing that the log-normal, besides providing an
good fit, is a functional form that emerges again and again in
the analysis of distributions in natural systems. From the failure
rate of electronic components \citep{sal08} to the latency periods of
infectious diseases \citep{kon77} and many other systems, the
log-normal distribution appears as the preferred rate model. Log-normal
distributions occur when multiplicative effects dominate; that the
cosmic SFRD is log-normal in time suggests a a deeper meaning than
that it is simply a good fit.

Figure 1 also shows fits to several other functional
forms suggested in the literature:
\begin{itemize}

\item The best fitting delayed exponential SFH, given by the
  $SFH\propto t/\tau^2 exp(-t^2/2\tau^2)$; the fit is obviously poor -
  with a $\chi^2$ of 10.2. This result is not unexpected, given the
  coupling of delay time and width in this model with a single shape
  parameter.

\item the best fitting exponential + power-law \citep{beh12}, given by
$SFH\propto t^A exp(-t/\tau)$; this model, with the same number of
parameters as the single log-normal, produces a poorer fit, with a
$\chi^2$ of 4.3.

\item the best fitting double power-law \citep{beh12}, given by
$SFH\propto ((t/\tau)^A+(t/\tau)^{-B})^{-1}$; this three-parameter
model fits the distribution extremely well, with a $\chi^2$ of 1.7.
\end{itemize}

The apparent agreement between different datasets seen in Figure 1
suggests that the uncertainties on the data as reported are
overestimated - or include a significant correlated systematic
component. We caution that a reduced $\chi^2$ is thus not trivial to
compute; however the relative $\chi^2$ of models with the same number
of parameters - i.e. the single log-normal and the exponential +
power-law is still informative. In this comparison the log-normal is
clearly preferred. 

Figure 1 also shows the result of a double log-normal fit (where each
log-normal component has equal weight). This fit resolves any
lingering tension between the peak height and late-time SFRD present
in the single log-normal fit. The $\chi^2$ of this fit is 1.6. We
include the double log-normal fit here primarily because it yields an
interesting physical interpretation; one of the fitted SFHs is
characterised as an early onset with a fast decline and the other as a
somewhat later onset with a slower decline. These two basic SFHs could
be interpreted as corresponding to early- and late-type galaxies, or
even the bulge and disk components of the typical galaxy. We also use
the double log-normal fit as a smooth description of the cosmic SFRD.

The simple model fits above have been computed using a downhill
simplex, but in general in this paper we fit ensembles of
parameterised SFH models to data using a simulated annealing
algorithm; the choice of simulated annealing to thoroughly explore the
parameter space for large aggregate samples of SFHs will
become apparent in the next section.

\subsection{The Specific Star Formation Rate at z=0}

The z=0 sSFR of a given galaxy is the current star formation rate
divided by total mass, and the total mass is the integral of the SFH
for that galaxy, adjusted for stellar mass loss. The z=0 sSFR
distribution is thus a reasonably orthogonal measure to the SFRD over
cosmic time, and a potentially useful contraint on a SFH model. We 
illustrate this point in Figure 2, which shows the $\tau$,~~$t_0$
parameter plane for a log-normal SFH, with lines of constant z=0
sSFR and lines of constant time of peak star formation shown. The
displayed range of sSFRs brackets values seen in several samples of
galaxies discussed below. Over much of that range, lines of
constant sSFR and lines of constant peak time are approximately
orthogonal --- i.e., imposing both a history and a current sSFR selects
one particular curve for most galaxies. Only galaxies with large sSFRs
at the present epoch are poorly constrained in this way.

As discussed in Paper III, an evolution in the rate of starbursts
in galaxies is insufficient to explain the observed evolution in the
sSFR distributions from low to high redshift. The presence of a main
sequence of star formation at both low and high redshifts has led
numerous authors \citep[e.g.,][]{noe07,rod11} to conclude that
starbursts are subdominant in affecting the observed evolution in star
formation in galaxies, as is also seen in simulations
\citep{dim08}. While it is apparent that secondary starbursts do
happen, our aim in this paper is to explore the ability - or lack
thereof - of smooth SFH models to reproduce the observed trends in
sSFR, completely absent any starbursts.

Motivated by Figure 2, we consider the z$\sim$0 sSFR distribution for
galaxies, for which we use the local galaxy sample described in Paper
III. Two sub-samples are included. The first, taken from the PG2MC
survey \citep{cal11}, covers a larger volume, but has a higher mass
limit of $4\times10^{10} M_\sun$, with minimum and maximum redshifts
of 0.03 and 0.11 and a median redshift of 0.0918. The second, from
SDSS observations of the northern galactic cap, is more restricted in
redshift, with minimum and maximum redshifts of 0.035 and 0.045 and a
median redshift of 0.0401. This second sub-sample has a lower mass
limit of $1\times10^{10} M_\sun$, and is cut at the upper end at the
lower limit of the first sub-sample. Galaxies are weighted to bring
the two sub-samples to a common volume. As described in \cite{oem12a}
(Paper I of this sequence) masses for this second sub-sample have been
computed using a variant of the technique in \cite{bell01} with
delayed exponential models (see \S2.1) as the underlying form of the
SFH. 

The total sample is 2094 galaxies, distributed in sSFR versus mass as
shown in Figure 3, with a mean weighted redshift of 0.0678.  The sSFR
values are computed from H$\alpha$ fluxes. The detectability of star
formation depends on H$\alpha$ equivalent widths and as a result the
sSFR limit is not trivial to describe, as it relies both on the flux
of the H$\alpha$ line in emission, as well as the continuum strength,
including the depth of the H$\alpha$ line in absorption. The resulting
incompleteness does not appear exactly fixed in either star formation
rate (this would be the expected result if only the H$\alpha$ line
flux were relevant, for example) or sSFR (which would be expected if
only the H$\alpha$ equivalent width were relevant). Figure 3 also
shows the fraction of galaxies which are measured as having a sSFR
identically zero as a function of mass; as expected early-type systems
with no measurable H$\alpha$ emission are proportionately more common
at the high mass end. For the purposes of this paper, we estimate the
threshold sSFR --- i.e., the allowed upper limit of the actual value
of the sSFR for a galaxy of a given mass measured in the data in Paper
III to have sSFR=0 --- as simply a fixed star formation rate of 0.05
$M_\sun yr^{-1}$.  Note that the exact choice of limits does not
significantly affect any of the results which follow.

To create a sample of SFHs that match both the $z\sim0$ sSFR
distribution, and the cosmic SFRD, we proceed as follows. We consider
a simulated sample of 2094 galaxies with masses identical to the sSFR
data discussed above. The SFH for each galaxy is described by two parameters, $\tau$
and $t_0$. We jointly solve for these parameters for each galaxy in
the ensemble using simulated annealing, requiring at the same time
that the mass- and sample-weighted sum of the individual SFHs match
the shape of the double log-normal model fit to the cosmic SFRD as detailed
in Figures 1 and 2.  We do not use the raw SFRD data detailed in
Figure 1, but the smooth fit to these data, since this modeling process
produces an under-constrained realisation rather than a unique
best-fit model.  Galaxies with a sSFR measured to be zero in the data
are allowed to take any value up to the mass-dependent threshold shown
in Figure 3.

The sSFR value for each simulated galaxy at its measured redshift is
computed simply as the ratio of the star formation rate divided by the
mass, where the latter is the integral of the former from early times
to the time of observation, modified downward by stellar mass loss that occurs
over the galaxy's history. We take the functional form of this mass
loss from \cite{jun01} and as in \cite{mar10} scale the mass loss so
that the total loss at 10 Gyr is 40\%. We do not attempt to compute the
mass loss individually for each galaxy, convolved across its star
formation history; rather, for computational simplicity and efficiency
we simply compute the mass loss from the peak of the SFRD at z$\sim2$
to the epoch of observation for each galaxy. The differences between
this approach and a more refined computation are in general small
because much of the mass loss occurs at early times, and by z$\sim$0
the bulk of the star formation in galaxies is well in the past.

The resulting distribution of sSFR values, and the distribution of
$\tau$ and $t_0$ parameters, is shown in Figure 4. The fit is
reasonable, however Figure 4 reveals a certain amount of tension
between the SFRD contraints and the z=0 sSFR contraints when fitting
each galaxy with a log-normal SFH. Specifically, the SFRD shape prefers
a somewhat higher normalisation of the z=0 sSFR distribution than the
actual measurements suggest. This is apparent in the fit shown in
Figure 4 in two ways: 1) the final fitted values of the sSFR for each
galaxy are on average, slightly high, and 2) the galaxies with sSFRs
that are nominally zero in the data are fit with sSFR values that
pile-up at the adopted upper threshold for the allowed sSFR
values. 

The source of this tension is, effectively, the mass limitation
of the input $z\sim0$ dataset, coupled to the changing contribution of
galaxies of various luminosities to the cosmic SFRD as a function of
redshift. \cite{cuc12} explore the contribution of galaxies of varying
far-ultraviolet luminosities and find that fainter galaxies contribute
more significantly at lower redshifts. This is the expected result
from an overall down-sizing in the star formation in galaxies, with
more massive (and generally more luminous) galaxies forming their
stars earlier, as first noted by \cite{cow96}. The consequence of this
effect in our analysis is that the contribution of the least massive
and faintest galaxies that are not included in our $z\sim0$ sample
varies as a function of redshift, and hence the nominal shape of the
cosmic SFRD as reported in \cite{cuc12} is not completely appropriate
to the particular set of $z\sim0$ galaxies we are considering here.


However, we might reasonably expect that any subset of galaxies
  drawn from the overall population would itself have a cosmic SFRD
  that is also well-described by a log-normal function in time. Indeed
  the SFRDs in \cite{cuc12} decomposed by luminosity appear to make a
  homologous set. Figure 5 demonstrates a model very similar to Figure
  4, except that we have allowed for an additional log-normal
  component designed to mimic the contribution of the faintest
  galaxies to the cosmic SFRD. We include this as a component whose
  mass is set at $5\times10^9M_\sun$ (a factor of two below the lower
  mass threshold of the data) and with an sSFR set from the mean
  mass-sSFR relation apparent in Figure 3. The total weight of this
  component (i.e., equivalent to setting the number of such galaxies)
  is such that it produces 20\% of the total star formation in the
  SRFD diagram.  This extra component --- designed to mimic the
  contribution of the faintest and least massive galaxies --- is fit
  as part of the modeling and behaves exactly as expected ; i.e.,
  interpreted as individual galaxies these are somewhat later-forming
  systems with a peak of star formation at z=1.0 which contributes
  most significantly to the SFRD at z=0. This component is shown in
  Figure 5. The addition of this component resolves the tension in the
  model noted above.

Finally, note that measured precisely from the model in Figure 5 at the
weighted mean redshift of the input sSFR sample, this component
produces 43\% of the star formation at that redshift. Figure 6 shows
the mass-sorted cumulative star formation in the input data, as well
as the nominal asymptote that this additional component suggests. This
extra amount of star formation appears perfectly reasonable. Note that
this agreement is at least somewhat arbitrary, as our choice of weight
for this extra component does affect (and was informed by) Figure 6.

\section{Comparison to Specific Star Formation Rates to z=1}

\subsection{An Unconstrained Model to z=1}

That the realisation of a galaxy population described above matches
both the SFRD as well as the z$\sim$0 sSFR distribution does not in
itself clearly validate the choice of a log-normal SFH for individual
galaxies. For example, one could in principle assemble the SFRD from a
large ensemble of galaxies and from a large range of functional forms
(simple tophat or Gaussian SFHs for example), and one can imagine
that at least some of these may very well reproduce both the SFRD and
the z$\sim$0 sSFR distribtions. 

To provide some insight on this point, we next compare the results of
the above log-normal realisation to the data from Paper III on galaxy
populations at intermediate redshifts; specifically we compare the
predicted sSFR distributions to the measured values. In this first
analysis we simply compare the predictions of the model to that data;
a model using the higher-z data as a constraint is presented in
\S3.2. Data are drawn from two sources, as described in Paper III: the
field sample from the IMACS Cluster Building Survey (ICBS), and the
All-wavelength Extended Groth Strip International Survey \citep[AEGIS:
][]{dav07}. SFRs in the ICBS are computed from a heirarchy of methods
calibrated against 24$\mu$m fluxes, with H$\alpha$ used where
available, then H$\beta$, or [OII]3727\AA\ when not. 24$\mu$m fluxes
are used for the brightest such objects (see Paper I and III for
further details). As described in Paper III, the SFRs in the AEGIS
sample have been recalibrated to our measurements and our calibration
by reconsidering their 24$\mu$m fluxes and also measuring and
calibrating emission lines in a set of DEEP2 spectra (upon which the AEGIS sample is
based) using our own tools. All SFRs at all redshift are thus
measured, as best possible, on the same calibrated system.

We compare the model predictions to data in four redshift
bins : $0.2<z<0.4$ and $0.4<z<0.6$ from the ICBS field galaxy sample,
and $0.6<z<0.8$ and $0.8<z<1.0$ from AEGIS. The mass limits, the
approximate sSFR limits for 100\% completeness, the total number of
galaxies and the fraction above the sSFR 100\% completeness limit in
each redshift bin are given in Table 1. Masses for the ICBS samples
are as described in Paper I, computed in a manner akin to
\cite{bell01} but with delayed exponential SFHs as the underlying star
formation model. Masses for the AEGIS subsamples are taken from
\cite{dav07} but converted to a Saltpeter IMF to match the other
samples. 

\begin{deluxetable}{llcccc}
\tablecolumns{6}
\tablewidth{0pc}
\tablecaption{Field Galaxy Data at Intermediate Redshifts}
\tablehead{
\colhead{Sample} & \colhead{Redshift Range}& \colhead{Mass Limit}& \colhead{sSFR Limit}& \colhead{\# of Galaxies}& \colhead{Fraction above}\\
\colhead{      } & \colhead{              }& \colhead{(M$_\sun$)}& \colhead{ (yr$^{-1}$)}& \colhead{}& \colhead{sSFR Limit}}
\startdata
ICBS & $0.2<z<0.4$ & $1\times10^{10} $ & $2\times10^{-11}$  & 466 & 0.534 \\
ICBS & $0.4<z<0.6$ & $4\times10^{10} $ & $4\times10^{-11}$  & 296 & 0.405 \\
AEGIS & $0.6<z<0.8$ & $1\times10^{10} $ & $6\times10^{-11}$  & 843 & 0.604 \\
AEGIS & $0.8<z<1.0$ & $2\times10^{10} $ & $5\times10^{-11}$  & 706 & 0.704 \\
\enddata
\end{deluxetable}

Before comparing the log-normal modeled sSFR distributions to the
galaxy data two further corrections are necessary. First, note that
the modeling as constructed does not explicitly account for merging;
the individual star formation histories tagged to individual galaxies
at $z\sim0$ must be unmerged as one goes to higher redshift, since local
galaxies are built up, to some extent, from smaller components over
cosmic time. We will compare the model realisation to the actual data
by integrating the model over mass down to the appropriate mass limit,
so a simple treatment of the merger history of the model galaxies is
sufficient.  Specifically we wish to know whether any individual
galaxy in the z$\sim$0 sample remains above the mass limit in a higher
redshift sample, or has `un-merged' and one or more of its un-merged
components have been removed from the higher redshift sample by virtue
of now falling individually below the mass limit. We draw a
description of the mass-dependent rate for major mergers from \cite{xu12}
and from this compute the merger rate over redshift for
each z$\sim$0 galaxy from the present epoch back to a redshift
assigned at random from the redshifts measured in the higher redshift
bin. There are no strong trends in mass or sSFR versus redshift in
these higher redshift datasets, so this simple process should
adequately model the higher redshift data, and intra-bin variations in
merger rates and sSFRs, without overly complicating this computation
of a best-fit realisation.  Using this merger rate from Xu et al. we
construct many realisations of the merger history of each galaxy,
allowing for mergers, if they occur, with mass ratios up to a 3:1
split, and allowing mergers up to two steps deep in the merger tree
for each z$\sim$0 galaxy. From this ensemble of mock histories we
compute the likelihood that each z$\sim$0 galaxy (or a portion
thereof) remains in the sample in the higher redshift bin when cut at
the appropriate mass limit, and use these likelihoods as weights when
computing the final sSFR distributions for the model. Note the
correction due to accounting for merging in this way is not dramatic;
in the highest redshift bin the variation in weights across the entire
mass range is less than a factor of two.

The second correction is straightforward; when assessing a given SFH
at higher redshifts (closer in time to the peak of the star formation
in a given galaxy) somewhat less stellar mass has been lost.
We adjust the galaxy masses and sSFRs appropriately.

Figure 7 shows the comparison of the sSFR distributions measured from
the data reported in Table 1 and for the model realisation shown in
Figure 5. Modeled galaxies are assessed at the same redshift in the
higher redshift bin used to compute the merger histories. The
agreement is remarkably good. That the overall scaling matches ---
i.e., that the model tracks the movement of the sSFR distributions to
higher values at earlier times is not in itself surprising, as this is
implicit in the fact that the model matches the cosmic SFRD. However,
the fact that the shapes of the distributions at sSFRs higher than the
100\% completeness limit agree rather well with the measured
distributions - in all four redshift bins, comprising some 60\% of the
total timeline of star formation in the universe - is striking. The
agreement is not perfect - for example in the highest redshift bin the
model somewhat overpredicts the abundance of the highest sSFR galaxies
- but nevertheless it suggests quite strongly that the choice of a
log-normal SFH for each galaxy is well motivated.  This agreement need
not have happened; it is not explicit in the modeling process, as we
will explore further in \S5 below.

\subsection{A Constrainted Model to z=1}

The next step in our analysis is to create a model realisation that
uses the higher redshift sSFR data as an explicit constraint, in
addition to the z$\sim0$ sSFRs. In this realisation we retain the mass
distribution given by the z$\sim$0 galaxies. An explicit galaxy-by
galaxy comparison to the higher redshift data is not possible, since
the different datasets described in Paper III sample different
co-moving volumes to different mass and sSFR limits in each redshift
bin, and moreover a methodology to account in detail for the merging
of galaxies across redshift bins - beyond the simple calculation
sketched above - is not apparent. Thus any constraint must, to some
extent, integrate across mass to ameliorate differences in the galaxy
mass functions between the various datasets, and then compare
distributions of sSFR values or related quantities, rather than making
a galaxy-by-galaxy comparison. Indeed the comparison of model outputs
to sSFR distributions in \S3.1 does exactly that, integrating across
mass directly, down to the mass limit of each redshift bin, and then
simply comparing histograms of the logarithm of the sSFRs. This is
likely not an optimal approach for constraining the model, however.

A closer look at the sSFR versus mass distribution for each redshift
bin of the data, shown in Figure 8, illustrates the model-to-data
comparison we have chosen to impose as a constraint on the
model. Specifically, note that the sSFR versus mass values form a
linear sequence in log-log space: this is the star-formation main
sequence described by many authors at many redshifts \citep[see
][ for a recent sumarry]{rod11}. This sequence appears to have an
approximately constant width with mass. Over the redshift interval
considered here, there is also no strong evidence for an evolution in
the slope of this linear sequence; the zeropoint clearly evolves to
higher values of sSFR at higher redshift but the slope remains
approximately constant. To construct a distribution of sSFR-like
values to constrain the model we thus consider the cumulative
distribution of the difference between log(sSFR) at various redshifts
and this relation at z=0, over the appropriate mass range in each
redshift bin. The resulting cumulative distributions for the data are
given in Figure 8.

As in \S3.1 above, the effect of mergers is computed as a weight for
each z$\sim$0 galaxy, and appropriately used when computing the
cumulative distribution of residual specific star formation rates in
the model. This final realisation of the model, now additionally using
the sSFR data in the higher redshift bins as a constraint, is shown in
Figure 9.

We remind the reader that all galaxies in these final model
  realisations have only smooth SFHs described by log-normal
  functions. No starburst activity is included. Starbursts might
  induce rapid - albeit temporary - changes in a given galaxy's sSFR,
  and we might expect to see this reflected as a tension between the
  sSFR distributions at low and high redshift, when attempting to
  connect the low and high redshift datasets using a smoothly varying
  SFH. However, the model in Figure 9 successfully incorperates all of
  the sSFR distributions to z=1 without including any starbursts, a
  point discussed at length in the next section.

\section{The Star Formation Histories of Galaxies to z=1 and Beyond}

In Paper III we highlighted several important aspects of the star
formation history of galaxies required by the redshift evolution of
the sSFR distribution of intermediate redshift galaxies. In
particular, Paper III illustrates that even at these intermediate
redshifts some galaxies must be suprisingly young, having formed the
bulk of their stellar populations in their recent past, and moreover
that starbursts, or a change in the prevelance thereof, cannot expain
the enhancement of sSFR values seen towards higher redshifts. The
final model realisation above provides some further insight into the
age distribution.

First, consider Figure 10, which shows the cumulative sSFR fraction in
each redshift bin in the model, akin to Figure 4 in Paper III. Note
that the model does an excellent job of reproducing the main trends
seen in the data. This model has no starbursts, yet can reproduce the
observed data with great fidelity. This provides a companion datum to
previous statements regarding starbursts; not only can starbursts not
readily produce the observed evolution, we show here that a population
of galaxies with star formation histories that have $only$ a smooth
component in time can in principle produce the measured sSFR
distributions. This does not argue that starbursts do not change the
observed distributions of sSFR - they clearly must do so, for a small
subset of all galaxies observed at any epoch. However, the broad
trends in sSFR over a Hubble time can be reproduced completely absent a
starbursting component.

Figure 11 shows the modeled ages of galaxies across galaxy mass. Note
that, as suggested in Paper III, some galaxies are indeed young
overall, having formed only half their stars within the past few
Gyrs. Moreover there is a general trend towards lower masses for more
star formation at later times; the mean lookback time for galaxies at
2$\times10^{11}$ M$_\sun$ to have formed half their stars is about
$\sim$9 Gyr ago, with 90\% of the star formation typically completed
by $\sim$7 Gyr ago whereas for a 2$\times10^{10}$ M$_\odot$ galaxy
these numbers are $\sim$7 and $\sim$2 Gyrs ago respectively.

\section{The Significance of the Log-Normal SFH}

The ubiquity of log-normal distributions in nature, and the apparently
excellent fit of the log-normal distribution to the cosmic SFRD, and
its utility even as a description of the SFH of individual galaxies,
leads unavoidably to speculation on whether this distribution is
somehow special in the context of star formation, or more simply
provides a reasonable fit because it has the required basic
properties; i.e., it rises at early times, falls at lates times, and
provides a two-parameter family of curves with both a duration and
onset of star formation that are independent (unlike, for example, a
delayed exponential in which the rise time and peak time are described
by a single parameter).

To further explore this question, we next consider a model
constructed identically to that in \S2.2, except that the log-normal
SFH for each galaxy is replaced with a Gaussian SFH. This model has
the same number of degrees of freedom as the log-normal model
discussed in \S3.1. As before, the ensemble of 2094 $z\sim0$ galaxies
is fit jointly with the cosmic SFRD, producing a realisation that
matches both of these constraints. As in \S3.1 we then compute the model
sSFR distributions in the higher redshift intervals and compare to the
measured data; the comparison is shown in Figure 12, with identical
scaling as Figure 7 to facilitate comparison. The lack of agreement
achieved by this Gaussian SFH is striking - it cleary does not
describe the higher redshift galaxy population with any significant
fidelity, completely unlike the log-normal model. This implies that it
is not simply that the log-normal model contains two independent
parameters that provides the agreement seen in Figure 7, but rather
that the functional form is itself important.

As a further illustration of the efficacy of the log-normal star
formation history in describing most galaxies, consider Figure
13. This shows the computed star formation rate versus stellar mass
for the final model from \S3.2, but now evaluated over the redshift
interval $1.5<z<2.5$. Our aim here is to compare to the z$\sim2$ star
forming main-sequence from \cite{dad07}; the figure is formatted as
for Figure 1 from \cite{rod11} to facilitate comparison to that
updated work. As in \S3.1 we draw a redshift of evaluation for each
z$\sim$0 galaxy from the higher redshift interval considered; lacking
detailed information about the high- sample we draw redshifts with
equal probability from that interval. Also as in \S3.1 we have
computed the effect of mergers using merger rates from \cite{xu12};
in this particular application we allow mergers up to 3 deep into the
merger tree, so that any one z$\sim$0 galaxy can be broken into up to
eight components in the higher-z bin. As before, we do not attempt to
compute individual histories for each component; all un-merged pieces
of a z$\sim$0 galaxy have the same SFH. Nevertheless, the model data
shown in Figure 13 are a reasonable match to the z$\sim$2 star-forming
main sequence of \cite{dad07} --- despite no direct constraint being
applied to the model at these redshifts apart from the integrated
cosmic SFRD contraint. The fidelity with which this model can
reproduce this higher redshift data on star formation, when
constrained in detail only at lower redshifts, argues again that the
underlying log-normal SFHs are a close approximation to the actual
SFHs of individual galaxies.

Detailed analyses of the shape of SFH of galaxies, at both high and
low redshifts, from both observational data and theoretical modeling,
can also be described by a log-normal. For example, the recent
literature includes archeological efforts to measure the SFH of local
galaxies by resolved color-magnitude diagrams of their stellar
populations \citep[e.g.][]{wil11,dol02}. In a larger but still
reasonably local volume \cite{pan07} have analysed the SFH of galaxies
taken fom the Sloan Digital Sky Survey by a careful comparison of
spectra to spectral synthesis models. The aggregate analyses
\citep{pan07,wil11} effectively reproduce a SFRD that looks similar to
the data used in Figure 1 and elsewhere and so presumably can be fit
by a log-normal; interestingly the analysis of \cite{wil11} sees the
same elevated local SFRD from very low-mass galaxies that we were
forced to include in \S2.2. In Figure 14 we show a log-normal fit to
the measured cumulative star formation from \cite{pan07}; the fit is
excellent. This fit is distinct from the fit presented in Figure 1,
but of similar overall shape. Our goal here is not argue the exact
match or lack thereof between the analysis of \cite{pan07} and direct
in situ measurements of the cosmic SFRD but simply to note that the
SFH suggested by those independent local data can indeed also be well fit by
a log-normal SFH. For comparison Figure 14 also shows a fit using the
power law + exponential form of \cite{beh12} - as in the fit to the
cosmic SFRD (c.f., Figure 1) this functional form is a reasonable
description of the data, though it provides a poorer fit than a single
log-normal.

At redshifts earlier than the peak in the cosmic SFRD \cite{pap11}
present the SFH of galaxies with a fixed comoving number density of
$2\times10^{-4}$ Mpc$^{-3}$ and show that their SFR rises with time;
they argue for a power-law description of the SHF of these galaxies
over the interval $3<z<8$, although they note that an inverted-$\tau$ model
is statistically indistinguishable from the power-law fit. We have fit
the SFH data from \cite{pap11} using a log-normal SFH in addition and
find that it is also statistically indistinguishable from a power law
as well (and is formally a better fit).

Finally, Figure 15 shows one further illustration of the utility of
the log-normal SFH. We have fit the mean SHFs derived from
cosmological hydrodynamic simulations in Figure 1 of \cite{fin11}
using log-normals. The rising mean SFHs for each mass bin from
Finlator et al. are extremely well described by a log-normal function
in each case.

\section{Future Work and Conclusions}

The modeling effort presented above can be significantly expanded in
several ways. Though we defer such work to further papers, we note
here for reference some future tests which may be applied to the
log-normal SFH framework that we have proposed.

\begin{enumerate}
\item The computation of spectrophotometric properties across galaxy
type and redshift: Such calculations were one of the main drivers for
the development of stellar population modeling. For example,
the simplest measure (broadband galaxy colors) are strongly correlated to sSFRs;
absent dust, colors are driven by the bluer light of the current star
formation overlaid on the redder light of the SFH. The mapping between
observed color and star formation rate is however non-linear, and
instantaneous changes in the star formation rate of a given galaxy
take several Gigayears to be fully expressed in the galaxy's colors,
and so a comparison of predicted and observed color distributions may
provide additional model contraints, and indicate the prevelance of
processes outside the star formation regime considered here (e.g.,
rapid changes - bursts or truncations - in the star formation history
of individual galaxies).
 
\item The apparent plateau in the sSFRs of galaxies at redshifts
  beyond z$\sim$2: Comparing models to observed galaxies will require
  a more robust treatment of mergers than we have used in this initial
  paper, since the strong relation between sSFR and mass in galaxy
  populations requires that models and data are compared over
  precisely the same mass range. The model proposed here may have the
  appropriate properties to produce the observed trend however; in the
  model realizations above most galaxies at times earlier than the peak
  of the cosmic SFRD have smoothly increasing SFHs and will not have
  the extremely high sSFRs one would expect from alternate models
  which start at a high SFR and then decline (such as a declining tau
  model with a delayed start).
\end{enumerate}

We also envision a further refinement to the model - simply that of
modeling each galaxy as a pair of log normal distributions, with an
adjustable weight between the two components. Such a complication is
motivated by the typical morphological structure of most galaxies,
with an old spheroidal component and a disky component with a more
extend star formation history. Though this would more than double the
number of adjustable parameters in the resulting model realizations,
it would also allow allow for additional constraints to be brought to
bear. For example by requiring that one of the two components have a
sSFR at the current epoch that is essentially zero (i.e., have a SFH
appropriate for the spheroidal component of a galaxy, but with a
varying weight set by the model) the model would produce (and could be
constrained by) measurements of the bulge/disk ratio in galaxies.

Regardless, the efficacy with which log-normal SFHs reproduce the
ensemble sSFRs of observed galaxies across a broad range of redshift
and mass, and the utility of the log-normal form in describing
the SFHs of a number of specific observed and simulated galaxy
populations, demonstrates that at minimum these SFHs are a useful
addition to the stellar population modeling toolkit.  Moreover, the
apparent ability of log normal SFHs to predict the sSFR distributions
of distant galaxies from only sSFR data at lower redshifts, coupled to
the integral constraint of the cosmic SFRD, suggests a deeper
significance - namely that the log normal SFH is a close approximation
to the actual star formation histories of most galaxies. This
conclusion is strengthened by the inability of a normal (rather than
log-normal) SFH model to reproduce the sSFR data; the shape of the
distribution matters. Individual galaxies aside, the simple conclusion
that the cosmic SFRD is lognormal in time seems to have not been
previously recognised, and likely has a physical significance.

Finally, given the overall success of the log-normal SFH modeling,
this paper provides further support to several of the conclusions of
Paper III, namely that some galaxies, even at intermediate redshifts,
are surpisingly young, and that the enhanced sSFRs of these objects
are not due primarily to starbursts.  The model realizations above
show that the sSFR distributions to z=1 can be fit with SFHs that are
only smoothly varying, completely absent any bursting
component. 



\acknowledgments
MDG thanks the Research Corporation for support of this work through a
Cottrell Scholars award. We thank an anonymous referee for two
rigorous and eminently helpful readings of the manuscript; their
effort and advice improved the final paper significantly.



{\it Facilities:} \facility{}

\onecolumn
\begin{figure}
\epsscale{1.0}
\plotone{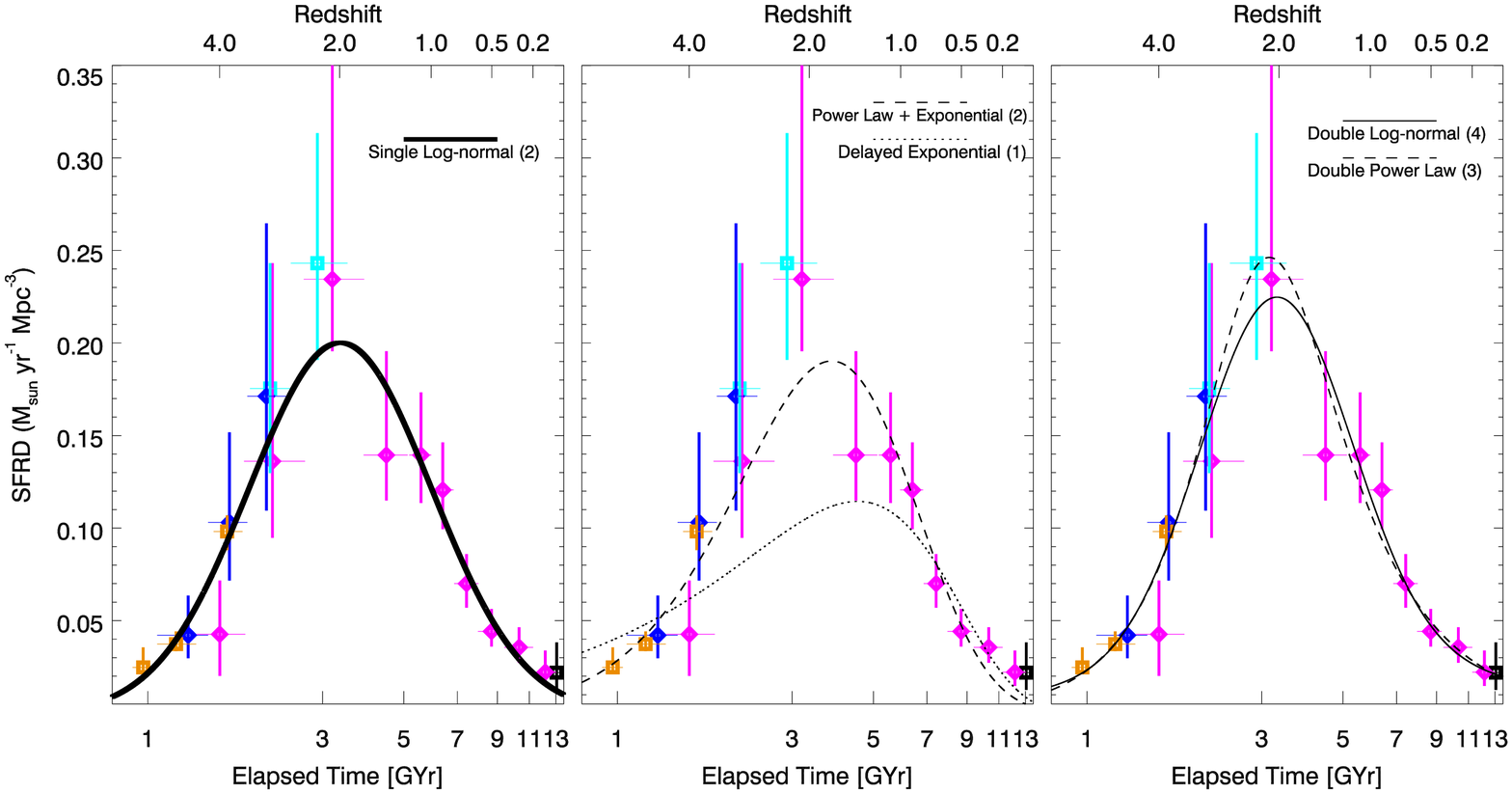}
\caption{SFRD measurements are from the literature, as compiled in
  \cite{cuc12}, plotted versus elapsed time. Magenta diamonds are the
  measurements of \cite{cuc12}; orange squares are from
  \cite{bou09}, blue diamonds from \cite{van10}, cyan squares from
  \cite{red09}, and the single low-z measurement in black is from
  \cite{wyd05}. Uncertainties are as reported in \cite{cuc12};
  horizontal lines indicate the redshift range over which each
  measurement was made. The left panel shows the best fit log-normal ($t_0$=1.539; $\tau$=0.574);
  this simple form appears to be an apt description of the cosmic
  SFRD.  The center panel shows the best fits using two SFH
  models from the literature - the powerlaw+exponential from \cite{beh12}
  and the delayed exponential \citep{gav02} - neither of which is a
  good fit to the data. The right panel shows the fit achieved with
  two more complex models - the double powerlaw from \cite{beh12} and
  a double log-normal ($t_0$=1.394,2.803; $\tau$=0.459,1.187). 
  In all panels the number of shape parameters in
  each fitting function is indicated parenthetically.
  \label{fig1}}
\end{figure}

\clearpage
\begin{figure}
\epsscale{1.0}
\plotone{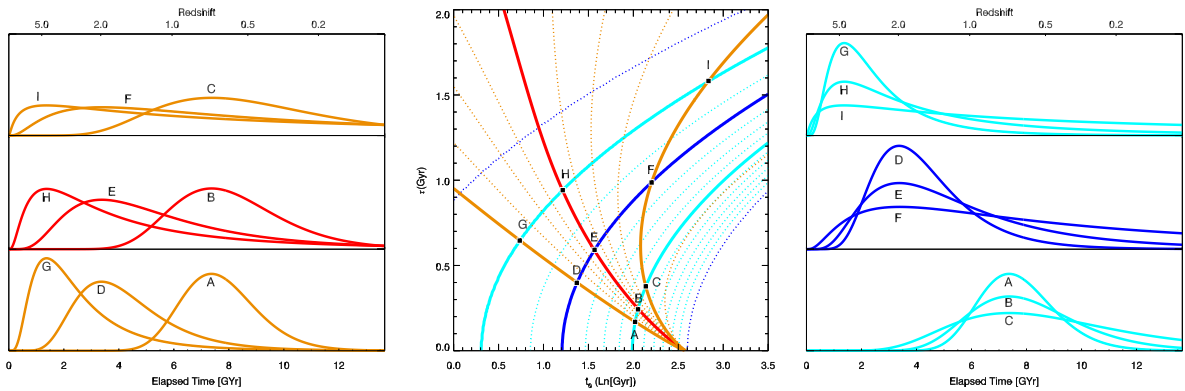}
\caption{{\it Center Panel:} Lines of constant z=0 sSFR (red,orange)
and constant time of peak star formation (blue,cyan) in the
$\tau$,$t_0$ parameter plane. Constant sSFR lines are shown at 5\%,
15\%...95\% of the non-zero sSFR rates in the dataset described later
in \S2.2. Galaxies with no measured star formation will appear to the
bottom left of this figure. Blue lines show the the location of
$\tau$,~$t_0$ parameter values for redshifts of 10.5, 2.0 and 0.0 (the
nominal epoch of reionization from WMAP \citep{lar11}, the peak of the
cosmic SFRD, and the present, respectively), with cyan lines spaced by
1 Gyr centered earlier and later than z=2.0. {\it Secondary Panels:}
individual log-normal SFHs are shown along the heavier lines in the
main panel, at the intersections of the heavy lines as picked out by
black point and labeled both in the main panel and in secondary
panels; panels to the right show log-normal distributions along lines
of constant time of peak star formation (for example, the middle of
these three panels, colored blue, shows SHFs which peak at z=2, along
the heavy blue line in the main panel) and panels to the left show
log-normal distributions along lines of constant z=0 sSFR (for
example, the middle of these three panels, colored red, shows SHFs
with a fixed z=0 sSFR along the heavy red line in the main panel). All
secondary panels are scaled to the same arbitrary peak SFR.
\label{fig2}}
\end{figure}

\clearpage
\begin{figure}
\epsscale{.60}
\plotone{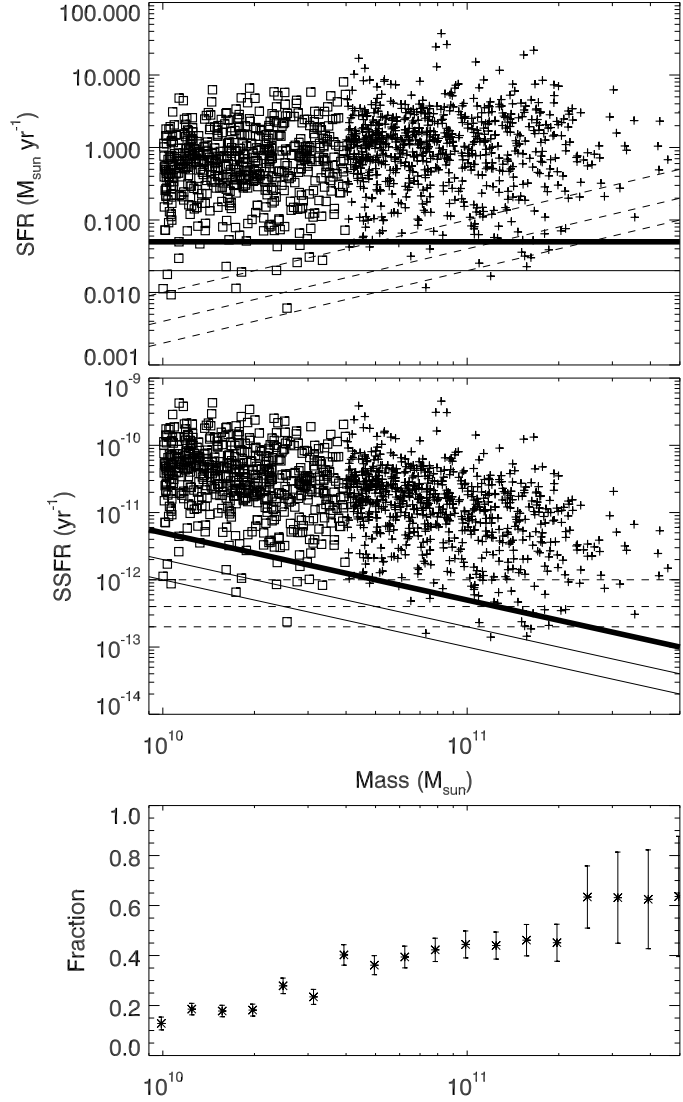}
\caption{The top two panels shows the distribution of star formation
  rates (top) and sSFRs (middle) versus mass at $z\sim0$ for the
  z$\sim$0 data from Paper III.  Objects reported as having a sSFR
  identically equal to zero are not shown. The data are drawn from two
  regions with differing volumes, and hence objects have differing
  relative weights in the combined sample (indicated by different
  symbols). Dashed lines are lines of constant sSFR (very
  approximately these are lines of constant H$\alpha$ equivalent
  width) at limits of $2\times10^{-13}$, $4\times10^{-13}$ and
  $1\times10^{-12} yr^{-1}$.  Solid lines are lines of constant star
  formation rate (effectively lines of constant H$\alpha$ flux) at
  limits of 0.01, 0.02 and 0.05 $M_\odot yr^{-1}$. The thicker solid
  line indicates the adopted threshold for galaxies with sSFRs
  identically equal to zero, as discussed in the main text. The bottom
  panel shows the fraction of these galaxies as a function of mass, in
  logarithmically spaced bins.
\label{fig3}}
\end{figure}

\clearpage
\begin{figure}
\epsscale{1.0}
\plotone{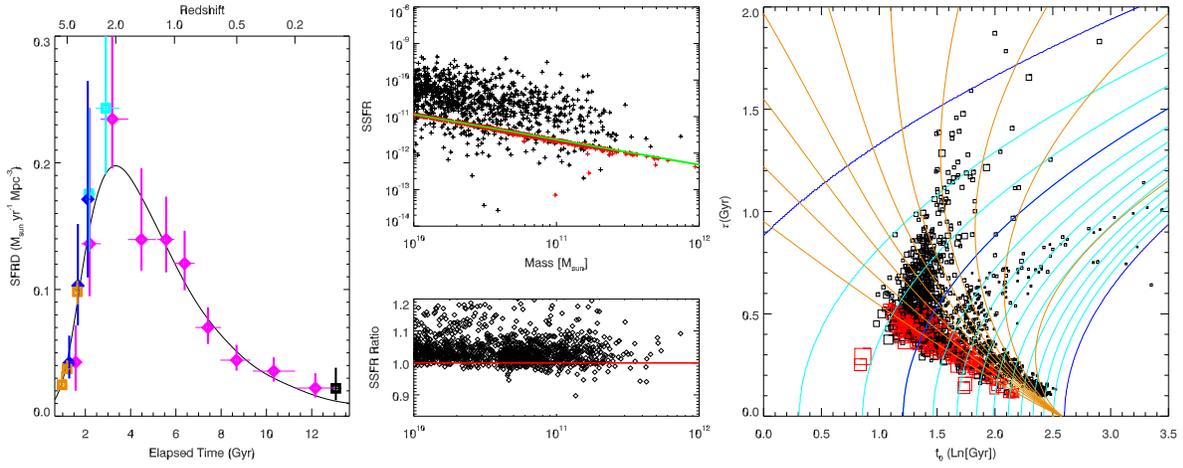}
\caption{{\it Left Panel:} The fit to the cosmic SFRD from a model
realisation which jointly fits these data as well as the z$\sim$0 sSFR
distribution. Data, symbols and colors are as in Figure 1. {\it Center
Panels, Top:} The corresponding fitted sSFR values for galaxies with a
measured non-zero sSFR (black) and those which are reported with a
sSFR of exactly zero (red). The allowed maximum value of the sSFR for
these nominally non-star-forming galaxies, described in Figure 3, is
shown by the green line. This model realization does not include any
allowance for missing low-mass but high-sSFR galaxies; note the
pile-up of nominally non-star-forming systems (the red symbols along a
single line) at the upper limit. {\it Center Panels, Bottom:} The ratio
of the fitted sSFR values to the measured sSFR values for the black
points (galaxies with measured non-zero sSFR) in the upper panel;
again note the tendency toward higher sSFRs in the model. {\it Right
Panel:} The distribution of log-normal parameters $t_0$ and
$\tau$. Symbols are individual galaxies with colors having the same
meaning as the top-center panel, and symbol size is proportional to
the galaxy mass.  The lines are as in Figure 2, though simplified.
\label{fig4}}
\end{figure}

\clearpage
\begin{figure}
\epsscale{1.0}
\plotone{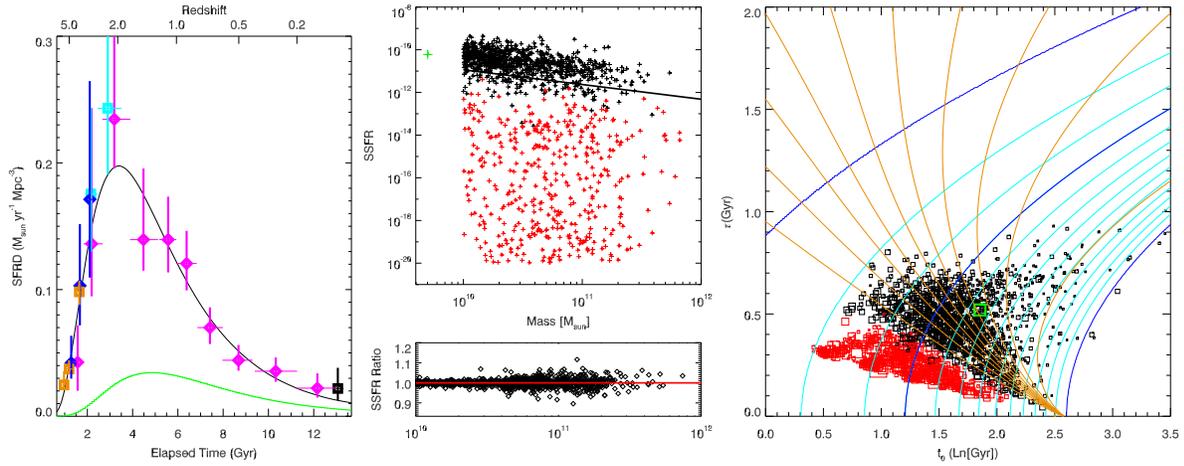}
\caption{Similar to Figure 4, except for the addition of a single
extra log-normal component which accounts for the low-mass but
high-sSFR galaxies not present in the z$\sim$0 sSFR data, and which
effectively resolves the tension between the cosmic SFRD and the
z$\sim$0 data.
\label{fig5}}
\end{figure}

\clearpage
\begin{figure}
\epsscale{1.0}
\plotone{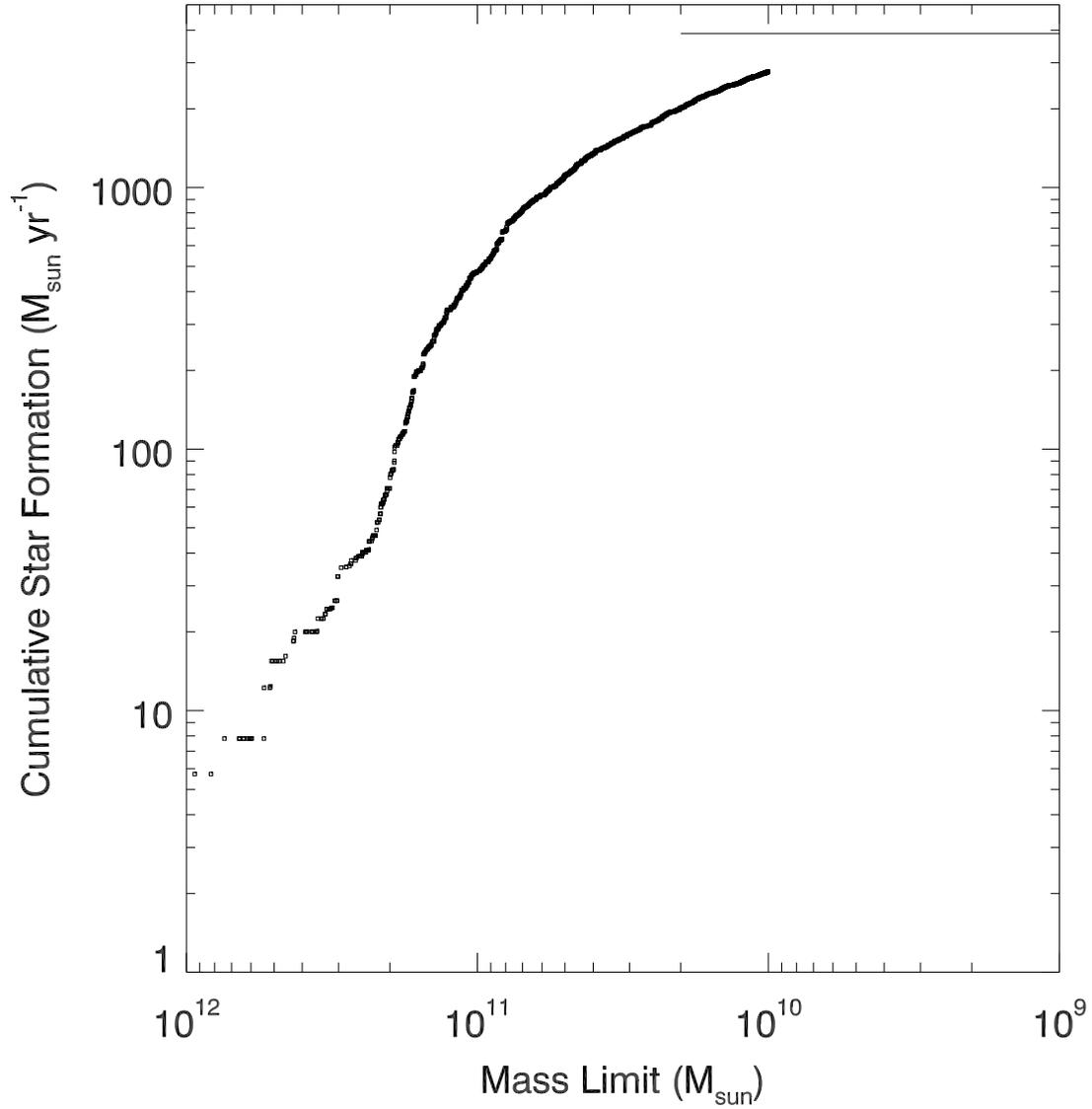}
\caption{The cumulative total star formation in the z$\sim$0 galaxy
sample, down to a given mass limit. At the lower mass limit it is
clear that not all of the local star formation is accounted for. The
horizontal line shows the implied total star formation in the sample
including the missing component included in the extra low-mass
log-normal component shown in detail in Figure 5; this total is a
resonable asypmtote given the measured data.  \label{fig6}}
\end{figure}

\clearpage
\begin{figure}
\epsscale{1.0}
\plotone{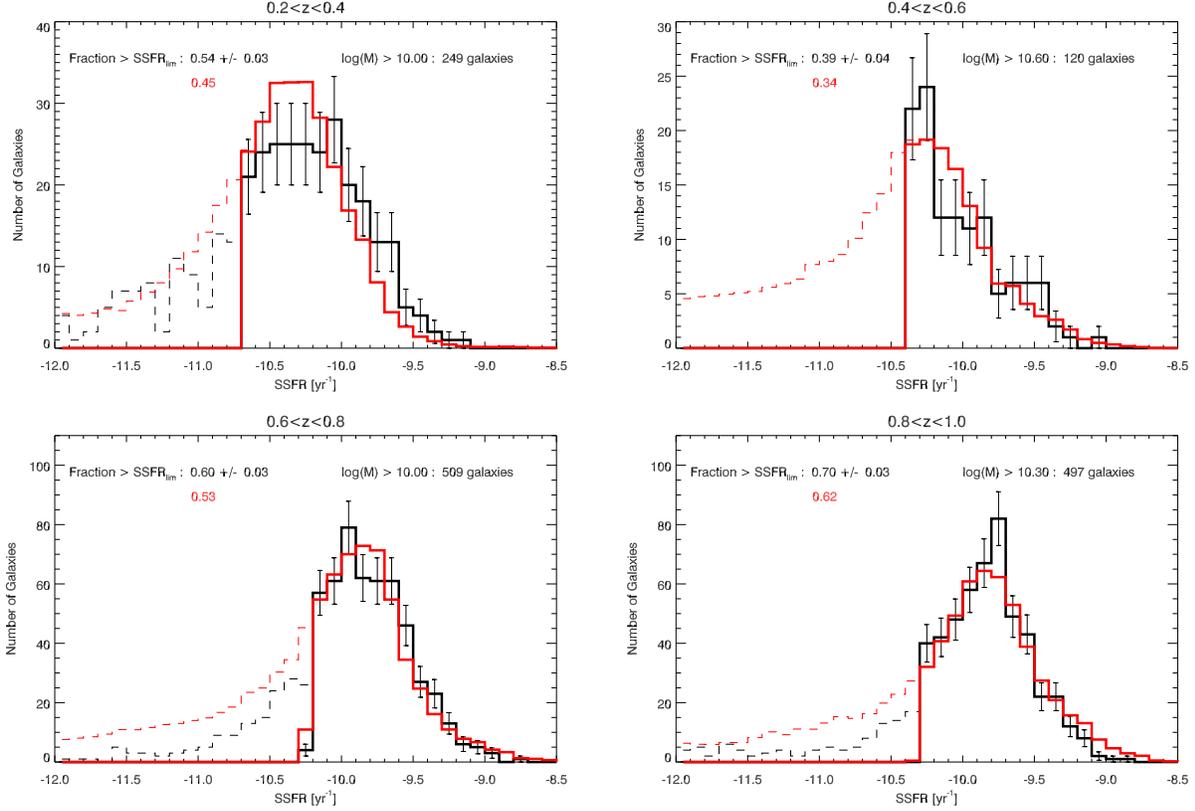}
\caption{The comparison between the measured sSFR distribution (black)
and the model prediction (red) for the model realisation shown in
detail in Figure 5, in four distinct redshift bins. Mass limits for
each redshift range are as indicated.  Thick lines show the sSFR range
thought to be complete; thin dashed lines show the region that is
incomplete. Many of the measured galaxies in this latter region have
sSFRs which are nominally zero, and so do not contribute to these
plots. The fraction of galaxies above the fiducial completeness sSFR
is shown in each panel for both the measured galaxies and the model
realisation. Uncertainties reported on the data are simple 1$\sigma$
counting errors. Given the somewhat arbitrary details of the treatment
of the galaxies with sSFRs nominally zero, and the uncertainty in the
actual sSFR completeness value, we anticipate that the actual
uncertainty on the reported fractions is a factor of several times
higher than reported and likely dominated by systematic effects. For
example, modifying the sSFR completeness limits by on order of 10\%
produces a change comparable to the quoted random uncertainty. Overall
the fit between the data and the model prediction is remarkably good.
\label{fig7}}
\end{figure} 

\clearpage
\begin{figure}
\epsscale{1.0}
\plotone{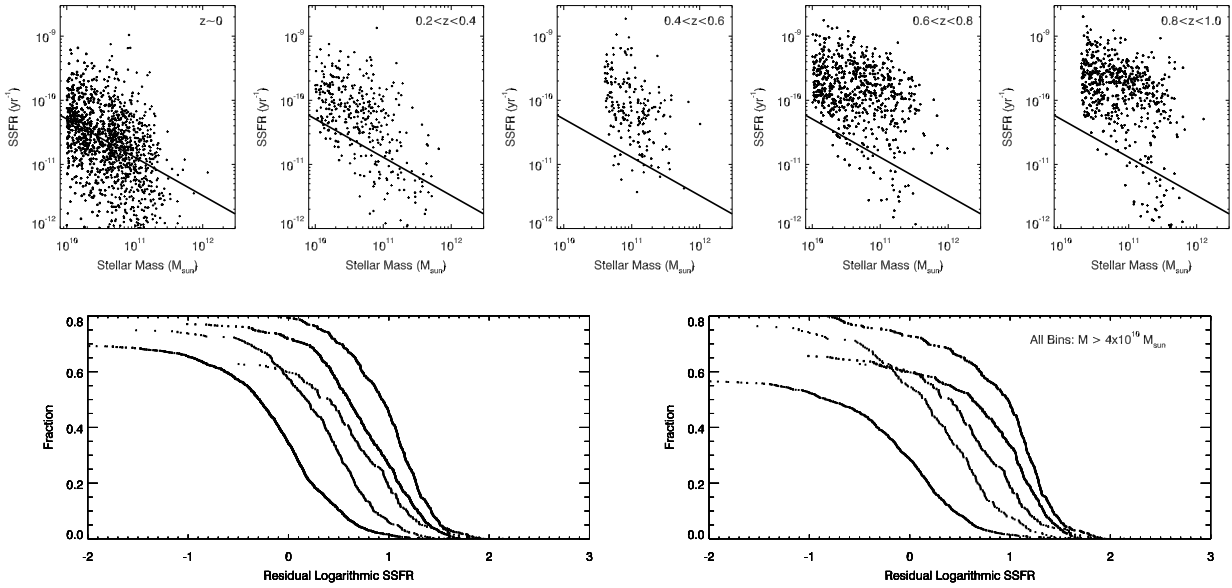}
\caption{{\it Top Panels:} sSFR versus mass for the data summarized in Table 1; redshift bins 
increase in redshift to the right from the z$\sim$0 sample on the
  left. The solid line shows the nominal z$\sim$0 star formation main
  sequence, fit using these data, and reproduced in the other
  panels for comparison. The two bottom panels show residual
  distributions of log(sSFR) minus this trend, computed to the
  limiting mass of each redshift bin (left panel) or to a common limit
  of $4\times10^{10}$ M$_\odot$  (right panel). The distributions
  on the bottom left are taken as contraints on the final model, down to the
  estimated completeness limit in sSFR in each redshift bin. Below
  this limit, the cumulative fraction in the fitted model is required
  to be at least as large as the measured (incomplete) fraction.
\label{fig8}}
\end{figure} 

\clearpage
\begin{figure}
\epsscale{1.0}
\plotone{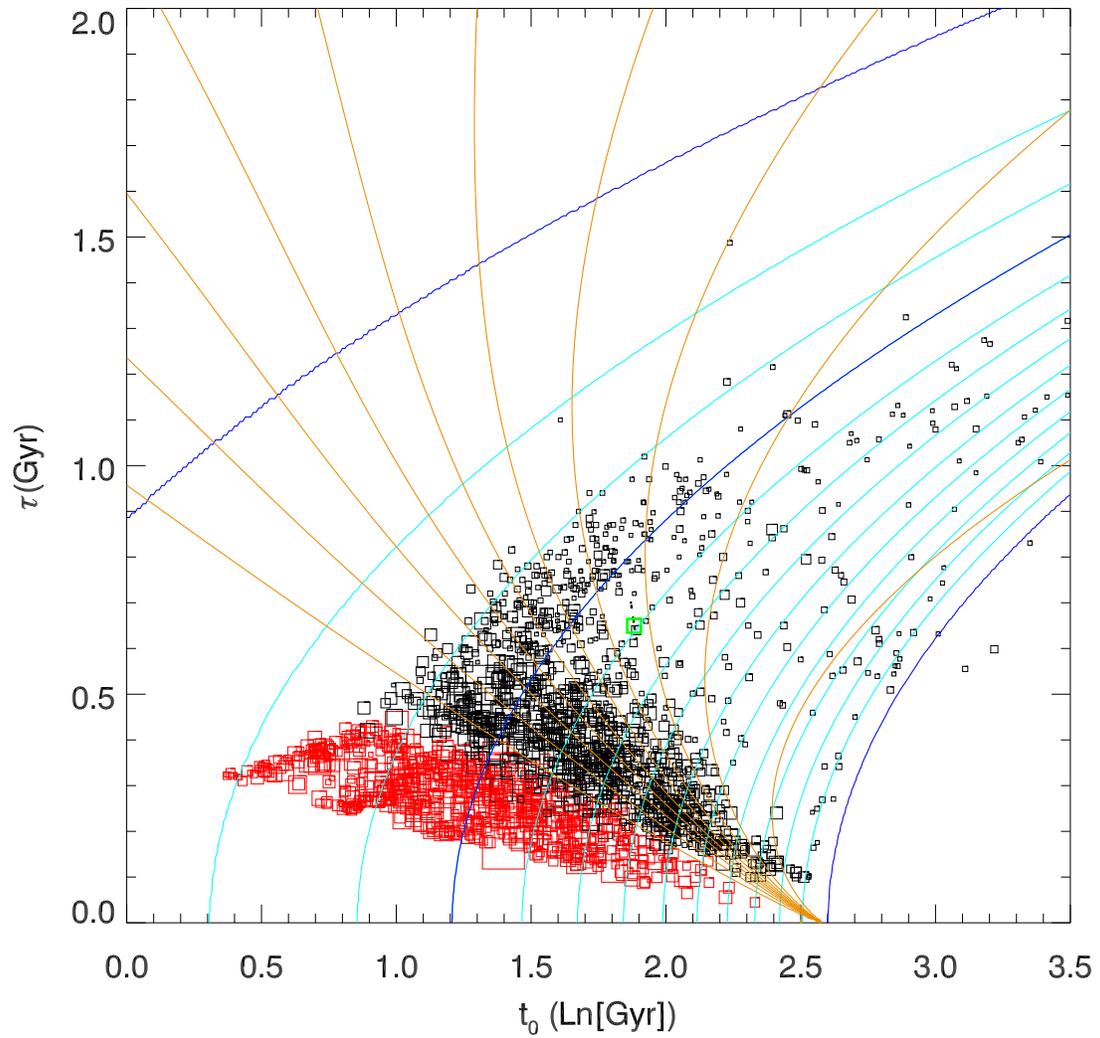}
\caption{The distribution of $t_0$ and $\tau$ parameters for the final
  model realisation, using the cumulative sSFR residuals as in Figure
  8 as constraints. Colors and symbols are as in Figures 4 and 5.
\label{fig9}}
\end{figure} 

\clearpage
\begin{figure}
\epsscale{1.0}
\plotone{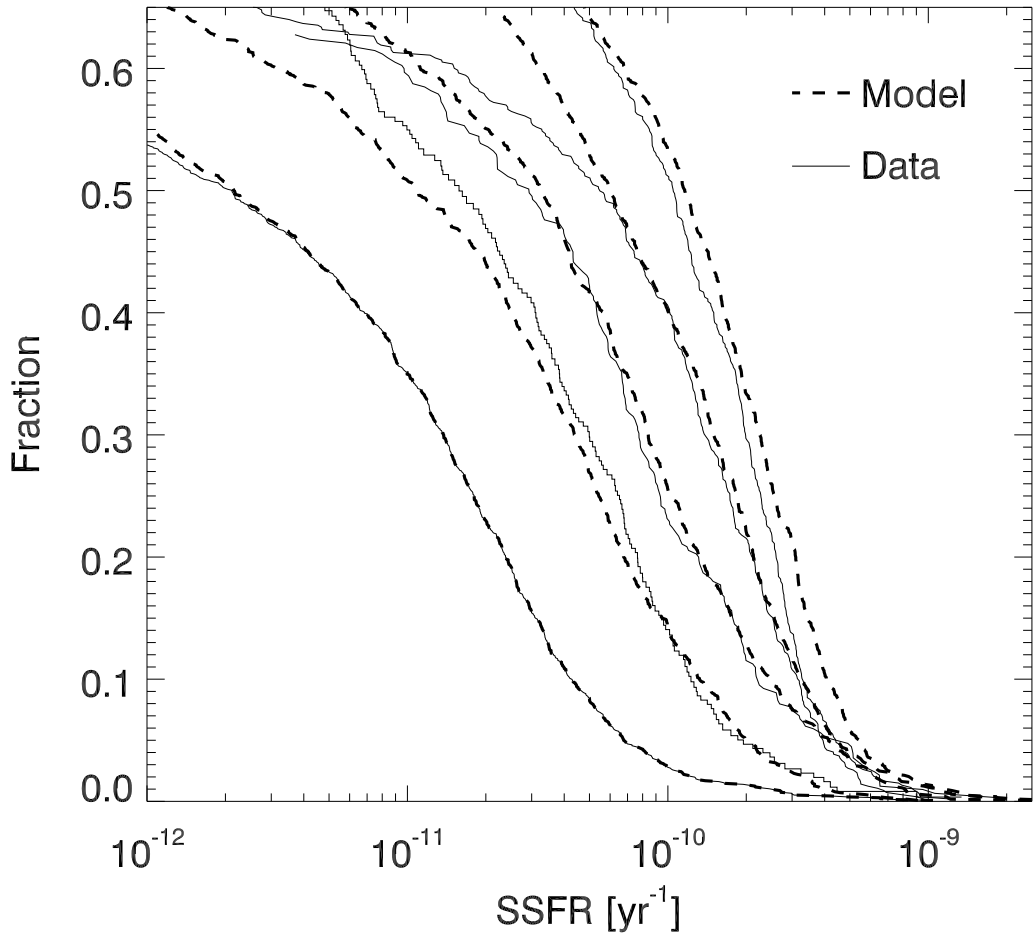}
\caption{The cumulative sSFR distribution for all five redshift
  intervals considered, limited to $M\geq4\times10^{10}$ M$_\odot$ as in
  Figure 4 of Paper III. Thin solid lines show the distributions from
  the data; heavier dashed lines are the results from the final model
  realisation dicussed in \S3.2. Redshift increases from left to right. 
\label{fig10}}
\end{figure}

\clearpage
\begin{figure}
\epsscale{1.0}
\plotone{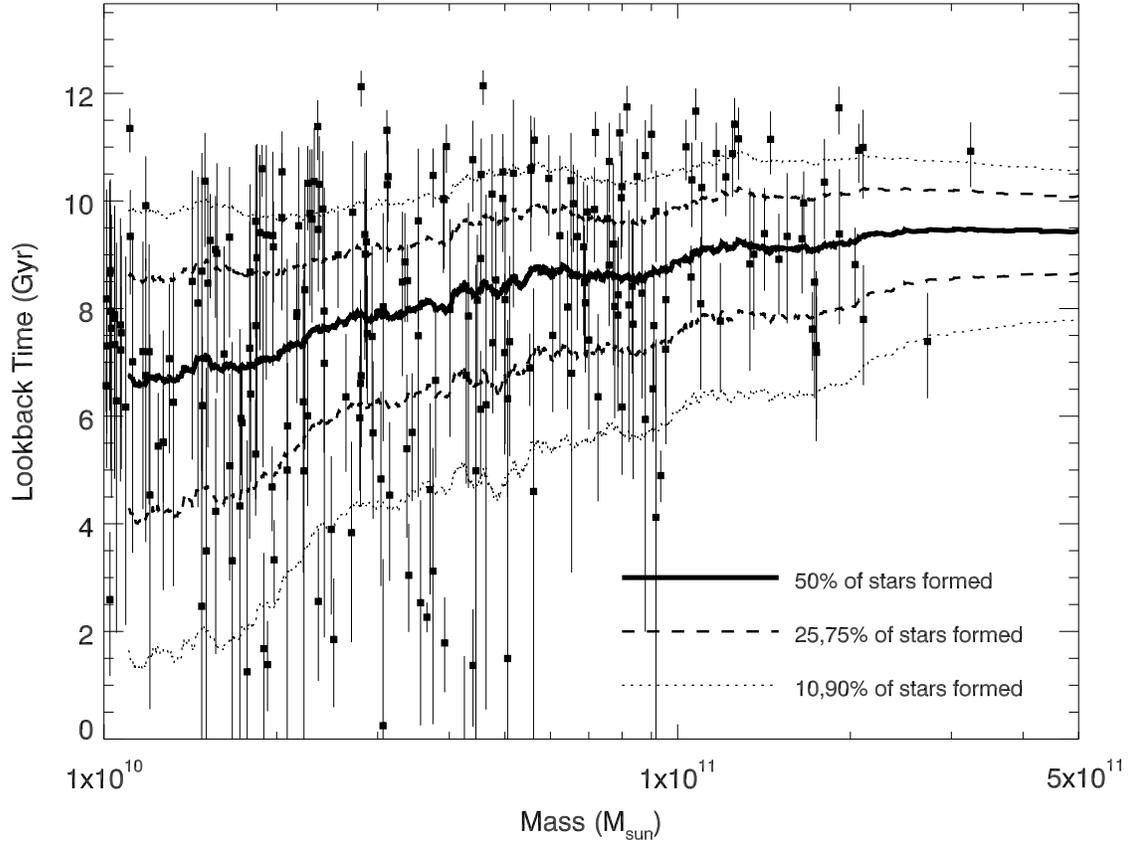}
\caption{The lookback time at which each modeled galaxy has completed
  50\% of its star formation (heavy symbols), for the final model
  using the higher redshift sSFR distributions as a constraint,
  against galaxy mass at $z\sim$0. Only some galaxies are plotted, for
  clarity. {\it Vertical lines show the time interval over which each
  galaxy forms 10\% to 90\% of its stars.} Overplotted curves show the
  typical value of the 10\% and 90\% times (thin dotted lines) the
  25\% and 75\% times (heavier dashed lines) and the 50\% time (heavy
  solid line). These typical values are computed as the mean in a
  running window across mass, 150 galaxies wide. Note that unlike most
  other times discussed elsewhere in this paper, this figure uses
  lookback time from z=0.
\label{fig11}}
\end{figure} 

\clearpage
\begin{figure}
\epsscale{1.0}
\plotone{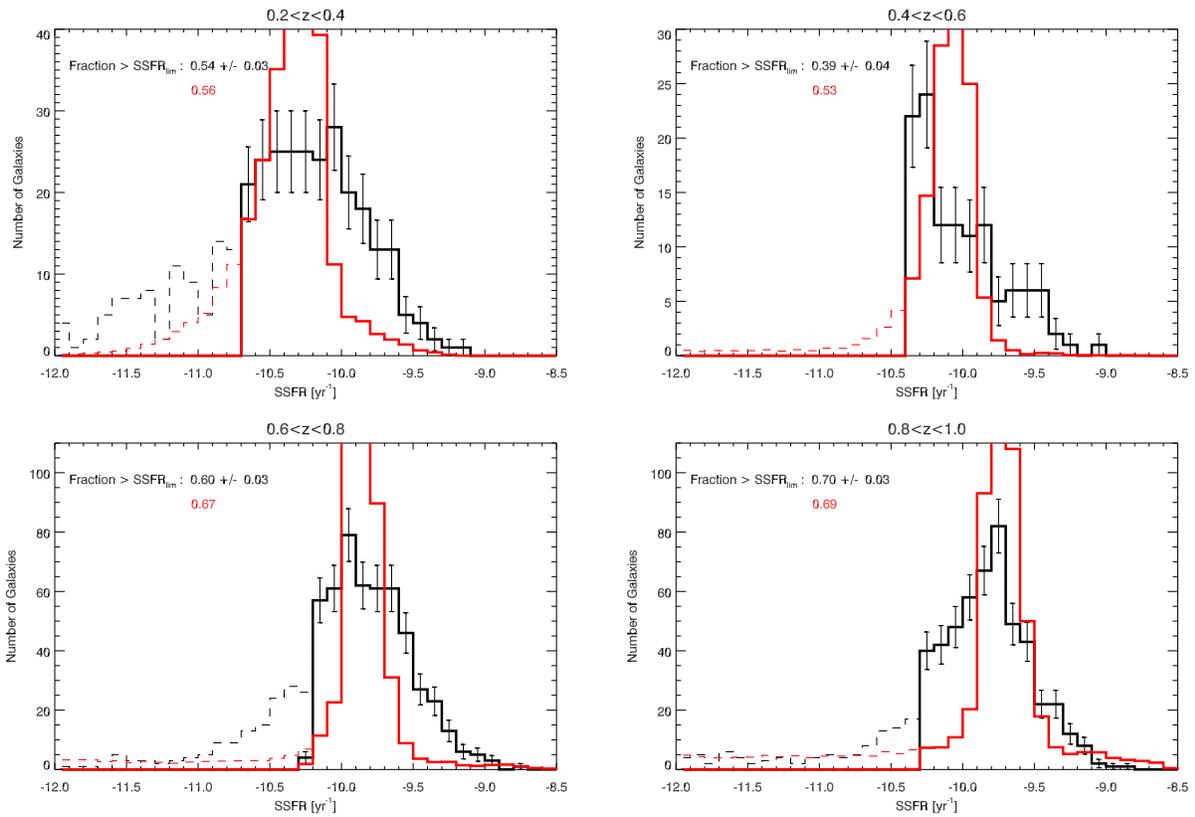}
\caption{As Figure 7, except that the model used has Gaussian SHFs for
galaxies rather than log-normal SFHs. The model is constrained by the
z$\sim$0 sSFR distribution and the cosmic SFRD as for the model shown
in Figures 5 and 7. The higher redshift predictions for this model are poor by
comparison to the log-normal model.
\label{fig12}}
\end{figure} 

\clearpage
\begin{figure}
\epsscale{1.0}
\plotone{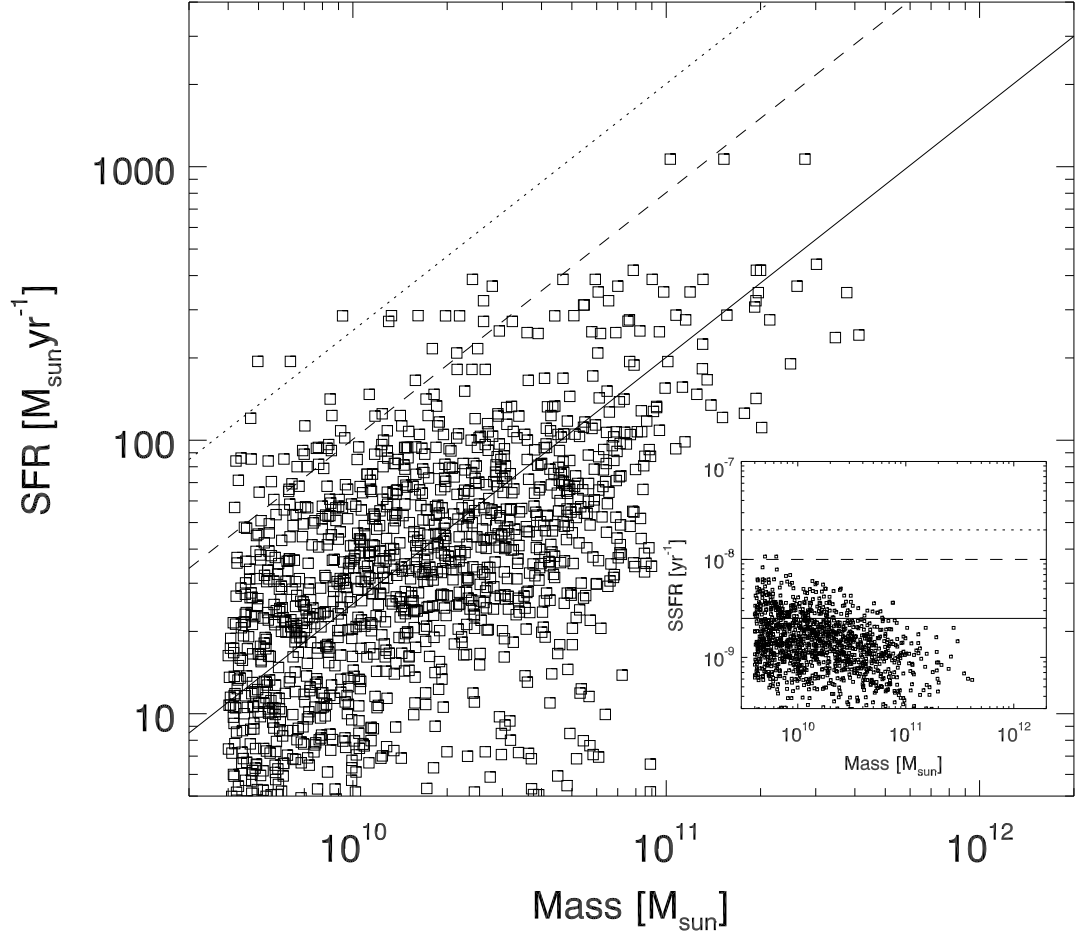}
\caption{Star formation rates versus stellar mass for the final model
(squares) evaluated in the redshift interval 1.5$<z<$2.5. The figure
is formatted as for Figure 1 from \cite{rod11} to
facilitate comparison to that updated work on the measured SFR-mass
distribution at z$\sim$2. The solid line in the main plot shows the
star-formation main sequence from \cite{dad07}, and the dashed
and dotted lines are 4 and 10$\times$ that relation (in SFR). As in
\cite{rod11} we also show the sSFR versus mass relation
for the same data in the inset plot; we use the same overplotted lines
as in that paper (these are $not$ exactly the main sequence line and
multiples shown in the main plot).
\label{fig13}}
\end{figure} 

\clearpage
\begin{figure}
\epsscale{1.0}
\plotone{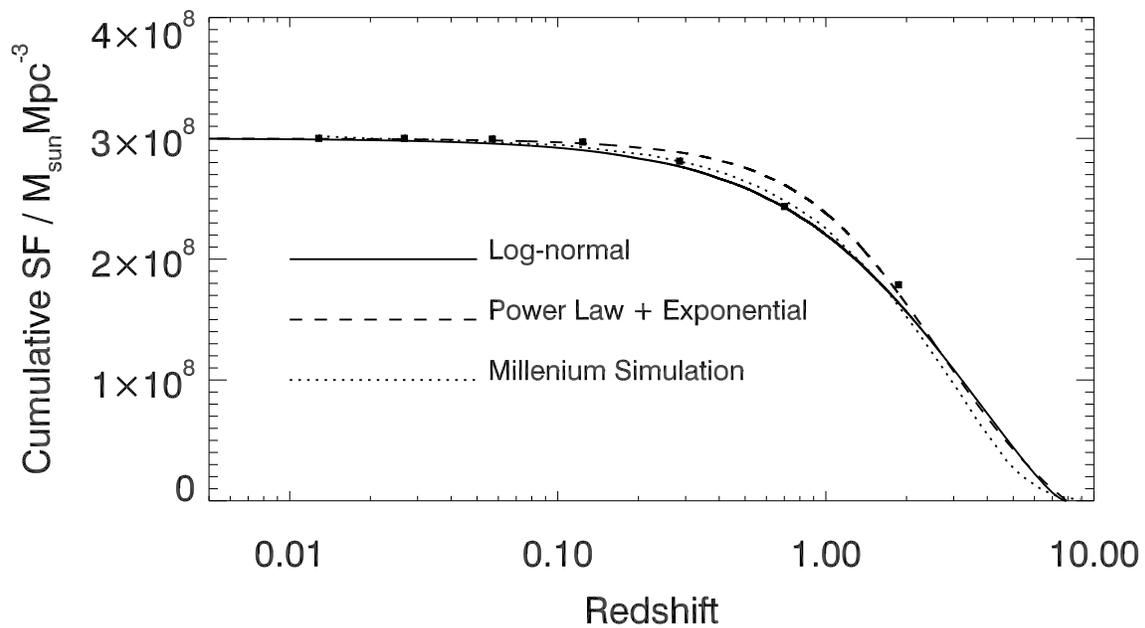}
\caption{The cumulative star formation versus redshift diagram from
  \cite{pan07}. Solid points show their measurements, and their scaled
  version of the predicted measurement from \cite{cro06} is shown as a
  dotted line. The solid line shows the best fit log-normal to these
  cumulative data, and the dashed line is the best fit power law +
  exponential. The best fit log-normal is given by $t_0$=1.564 and
  $\tau$=1.728.
\label{fig14}}
\end{figure} 

\begin{figure}
\epsscale{1.0}
\plotone{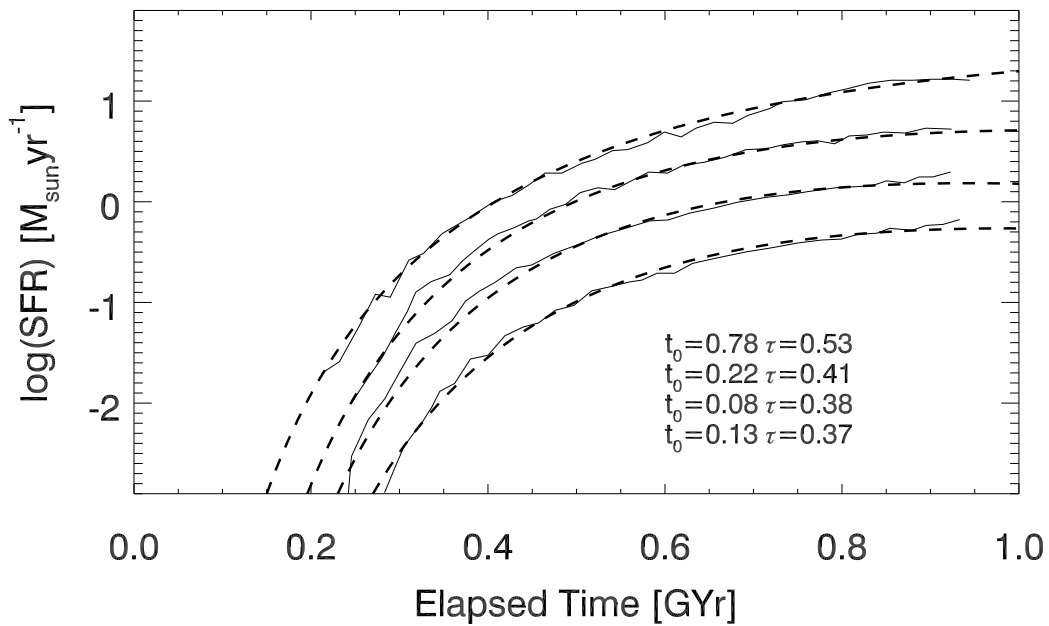}
\caption{The simulated mean SFH of galaxies in the early Universe
($z>5.5$) from \cite{fin11} (solid lines) in mass bins centered, from
top to bottom, at masses of $\log{M_*/M_\sun}=9.7,9.2,8.7,8.2$. The
dashed lines show best fitting log-normal function to each of these
mean SFHs.\label{fig15}} \end{figure}

\end{document}